\documentstyle[epsf,twocolumn,prc,aps]{revtex}

\begin{document}
\draft
\title{The effective $g_A$ in the $pf$-shell}
\author{G. Mart\'{\i}nez-Pinedo\thanks{gabriel@nuc2.ft.uam.es} and
  A. Poves\thanks{poves@nucphys1.ft.uam.es}} 
\address{Departamento de F\'{\i}sica Te\'orica C-XI,\\ Universidad
  Aut\'onoma de Madrid,\\ E--28049 Madrid, Spain}
\author{E. Caurier\thanks{caurier@crnhp4.in2p3.fr} and
  A. P. Zuker\thanks{zuker@crnhp4.in2p3.fr}} 
\address{Groupe de Physique Th\'eorique, Centre de Recherches
  Nucl\'eaires, Institut National de Physique Nucl\'eaire et de
  Physique des Particles, Centre National de la Recherche
  Scientifique, Universit\'e Louis Pasteur, Bo\^{\i}te Postale 28,
  F--67037 Strasbourg Cedex~2, France}
\date{\today}
\maketitle

\begin{abstract}
  We have calculated the Gamow-Teller matrix elements of 64 decays of
  nuclei in the mass range $A=41$--50. In all the cases the valence
  space of the full $pf$-shell is used. Agreement with the
  experimental results demands the introduction of an average
  quenching factor, $q=0.744\pm0.015$, slightly smaller but
  statistically compatible with the $sd$-shell value, thus indicating
  that the present number is close to the limit for large $A$.
\end{abstract}
\pacs{PACS number(s): 21.10.Pc, 25.40.Kv, 27.40.+z}

The observed Gamow Teller strength appears to be systematically
smaller than what is theoretically expected on the basis of the model
independent ``$3(N-Z)$'' sum rule. Much work has been devoted to the
subject in the last fifteen
years~\cite{gaarde,bertsch,osterfeld,goodman}. The heart of the
problem can be summed up by defining the reduced transition
probability as

\begin{equation}
  B(GT)=\left(\frac{g_A}{g_V}\right)^2 \langle
  \bbox{\sigma\tau}\rangle^2, \hspace{0.5cm} \langle
  \bbox{\sigma\tau}\rangle =\frac{\langle
    f||\sum_k\bbox{\sigma}^k \bbox{t}^k_\pm
    ||i\rangle}{\sqrt{2J_i+1}},
\end{equation}
and asking: Is the observed quenching due to a renormalization of the
${g_A}$ coupling constant ---originating in non nucleonic effects---
or is it the $\bbox{\sigma\tau}$ operator that should be renormalized
because of nuclear correlations?

The analysis of some $pf$-shell nuclei for which very precise data are
available and full $0\hbar\omega$ calculations are possible, strongly
suggests that most of the theoretically expected strength has been
observed~\cite{quenching,hierros} . The quenching factor necessary to
bring into agreement the calculated and measured values is directly
related to the amplitude of the $0\hbar\omega$ model space components
in the exact wave functions. This normalization factor can also be
obtained from $(d,p)$ or $(e,e'p)$ reactions and reflects the
reduction in the discontinuity at the Fermi surface in a normal
system. As such, it is a fundamental quantity, whose evolution with
mass number is of interest.

In principle there are two ways of extracting it from Gamow Teller
processes. One is to equate it to the fraction of strength seen in the
resonance region in $(p,n)$ reactions. The alternative is to calculate
lifetimes for individual $\beta$ decays and show that they correspond
to the experimental values within a constant factor. The latter
procedure is more precise, but demands high quality shell model
calculations that until recently were available only up to
$A=40$~\cite{quen-sd1,quen-sd2,quen-p}.

Our aim is to extend these analyses to the lower part of the $pf$
shell. Full $0\hbar \omega$ diagonalizations are done using the {\sc
  antoine} code~\cite{antoine} with the effective interaction KB3, a
minimally monopole modified version~\cite{KB3} of the original Kuo
Brown matrix elements~\cite{KB}. We refer to~\cite{a48} for details of
the shell model work.

Following ref.~\cite{towner} we define quenching as follows: for beta
decays populating well-defined isolated states in the daughter
nucleus, the square root of the ratio of the experimental measured
rate to the calculated rate in a full $0\hbar\omega$ calculation is
called the quenching factor. An average quenching factor, $q$, implies
an average over many transitions, and may be incorporated into
an effective axial vector coupling constant: 

\begin{equation}
  \label{q_def}
  q=\frac{g_{A,\text{eff}}}{g_A},
\end{equation}
where $g_A$ is the free-nucleon value of $-1.2599(25)$~\cite{towner}.
Following ref.~\cite{quen-sd1} we define

\begin{equation}
  \label{eq:mgt}
  M(GT)=\left[(2J_i +1)\,B(GT)\right]^{1/2},
\end{equation}
so as to have quantities independent of the direction of the
transition. Note here that our reduced matrix elements follow Racah's
convention~\cite{edmonds}.
In table~\ref{tab:mgt_comp} we list the $M(GT)$ values and compare
them with the experimental results. The table contain all the
transitions known experimentally. We also include the quantum numbers
of the final states, the $Q$-values, the branching ratios and the
experimental $\log ft$ values from which the $B(GT)$ values were
obtained using

\begin{equation}
  \label{eq:ft}
  (f_A+f^\epsilon)t = \frac{6146\pm6}{(f_V/f_A) B(F) + B(GT)}.
\end{equation}
the value $6.146\pm 6$ is obtained from the nine best-known
superallowed $0^+ \rightarrow 0^+$ decays~\cite{towner}. $f_V$ and
$f_A$ are the Fermi and Gamow-Teller phase-space factors,
respectively~\cite{wm,quen-sd1}. $f^\epsilon$ is the phase-space for
electron capture~\cite{bamby}.

A quick look to the table shows that the calculated values are
systematically larger than the experimental ones. In order to obtain
the effective $g_A$, first we  normalize the $M(GT)$ to the
``expected'' total strength, $W$ (listed in table~\ref{tab:mgt_comp})
and defined by

\begin{equation}
  \label{eq:w_def}
  W=\left\{\begin{array}{cc}
      \displaystyle{|g_A/g_V|\left[(2J_i+1)3|N_i-Z_i|\right]^{1/2}} & 
      \text{for } N_i \neq Z_i,\\[2mm] 
      \displaystyle{|g_A/g_V|\left[(2J_f+1)3|N_f-Z_f|\right]^{1/2}} &
      \text{for } 
      N_i = Z_i. 
    \end{array}\right.
\end{equation}
 
In figure~\ref{fig:rgt} are plotted the experimental values versus
the theoretical ones for
\begin{equation}
  \label{eq:rgt}
  R(GT)=M(GT)/W.
\end{equation}

The points follow nicely a straight line whose slope gives the average
quenching factor, $q=0.744\pm0.015$.  Most $R(GT)$ values are much
smaller than 1, reflecting the fact the strength in the decay window
is small and fragmented.  As a consequence, each individual decay may
be sensitive to small uncertainties in the calculations, which can be
averaged out by summing the total strength for each nucleus. Therefore
we introduce a new quantity

\begin{equation}
  \label{eq:tgt}
  T(GT)=\left[\sum_f R^2(GT, i \rightarrow f)\right]^{1/2}.
\end{equation}

In the corresponding plot in figure~\ref{fig:tgt} the points again
follow closely the $q=0.744$ line.

Comparing with the results in other regions, is suggestive
\begin{itemize}
\item $pf$ shell, $q= 0.744\pm0.015$ this work,
\item $sd$ shell, $q= 0.77\pm0.02$~\cite{quen-sd2},
\item $p$ shell, $q=0.82\pm0.015$~\cite{quen-p}.
\end{itemize}

In the figures, both the lines for $q=0.744$, and $q= 0.77$ are drawn
and it is clear that there is not much to choose between them, and
indeed, the average quenching factor of 0.77 has been extensively used
in many $pf$-shell calculations, either in direct
diagonalizations~\cite{a48,quenching,hierros} or Shell Model Monte
Carlo studies~\cite{smmc}, leading to agreement with global
Gamow-Teller strengths (as measured in $(n,p)$ and $(p,n)$ reactions)
and lifetimes.

Nevertheless, the results in the three regions point to a decrease of
$q$ with mass number, and the closeness of the $sd$ and $pf$ values
suggests that we have reached the large-A regime. This observation is
quite consistent with the numbers extracted by Osterfeld from $(p,n)$
data in heavier nuclei (see fig.~6 in~\cite{osterfeld}).

Whatever its origin, the $q$ factor is a fundamental quantity telling
us about correlations that are so well hidden, precisely behind
overall renormalizations such as $q$, that their existence may be
doubted.

This work is partially supported by DGICYT (Spain), Grant
No.  PB93-263, and by the IN2P3 (France) CICYT (Spain) collaboration
agreements.

\onecolumn

\begin{minipage}[h]{0.45\textwidth}
\begin{figure}
  \begin{center}
    \leavevmode
    \epsfxsize=\textwidth
    \epsffile{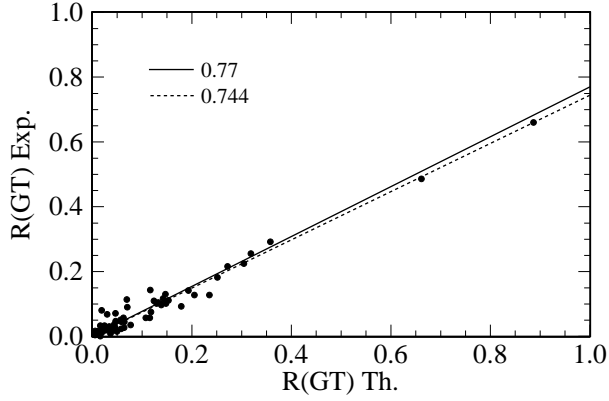}
    \caption{Comparison of the experimental matrix elements $R(GT)$
      with the theoretical calculations based on the ``free-nucleon''
      Gamow-Teller operator. Each transition is indicated by a point
      in the $x$-$y$ plane, with the theoretical value given by the
      $x$ coordinate of the point and the experimental value by the
      $y$ coordinate.}
    \label{fig:rgt}
  \end{center}
\end{figure} 
\end{minipage}
\hspace{0.5cm}
\begin{minipage}[h]{0.45\textwidth}
\begin{figure}
  \begin{center}
    \leavevmode
    \epsfxsize=\textwidth
    \epsffile{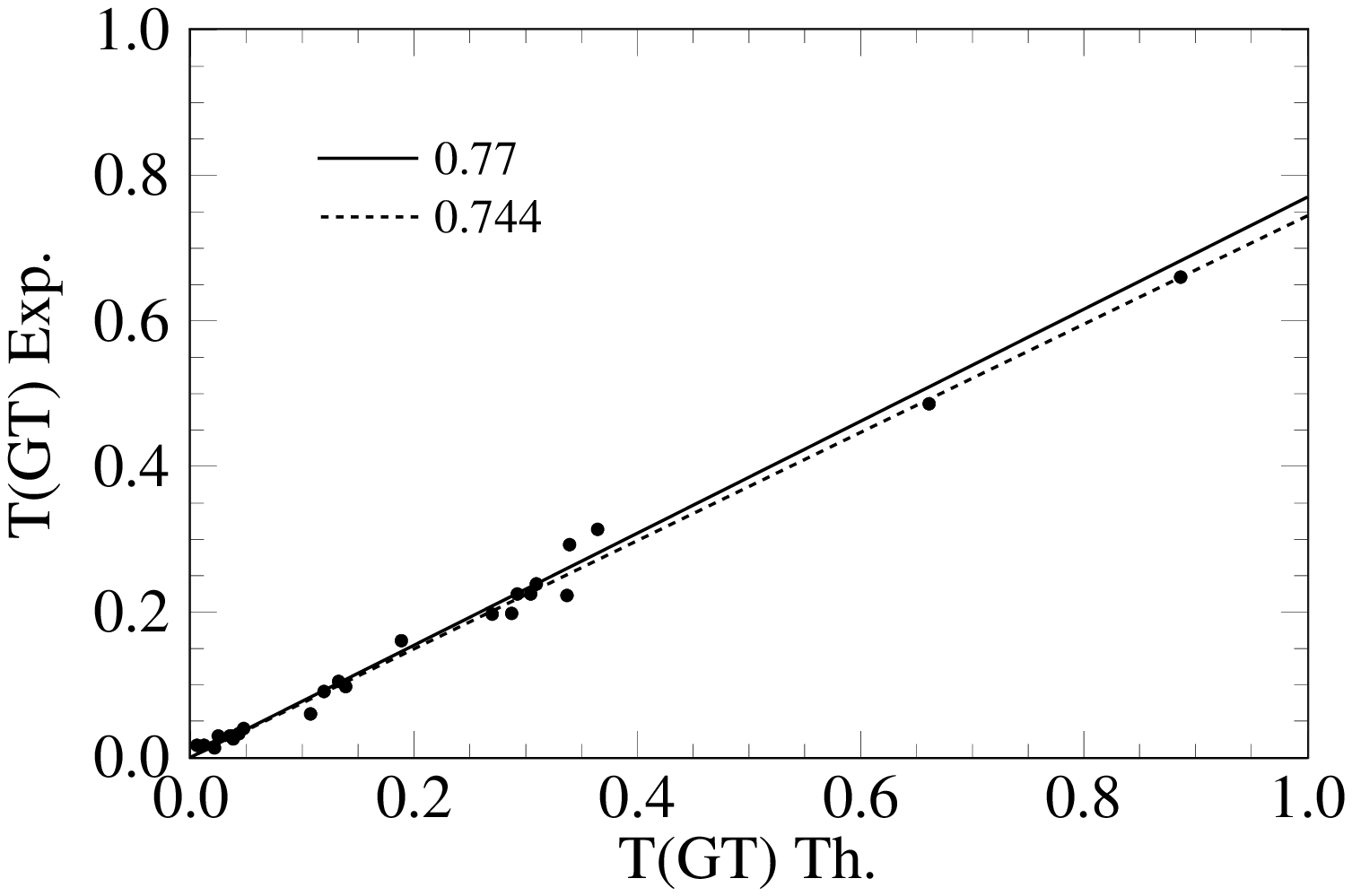}
    \caption{Comparison of the experimental values of the sums $T(GT)$
      with the correspondig theoretical value based on the
      ``free-nucleon'' Gamow-Teller operator. Each sum is indicated by
      a point in the $x$-$y$ plane, with the theoretical value given
      by the $x$ coordinate of the point and the experimental value by
      the $y$ coordinate.}
    \label{fig:tgt}
  \end{center}
\end{figure}
\end{minipage}

\begin{table}
  \begin{center}
    \leavevmode
    \caption{Experimental and theoretical $M(GT)$ matrix elements. The
      experimental data have been taken from~\protect\cite{nds}.
      $I_\beta + I_\epsilon$ are the branching ratios . All other
      quantities explained in the text.}
    \label{tab:mgt_comp} 
    \begin{tabular}{lccccrrr}
      Process  & $2J_n^\pi, 2T_n^\pi$ & $Q$  & $I_\beta +
      I_\epsilon$ & $\log ft$ & \multicolumn{2}{c}{$M(GT)$} & $W$ \\
      \cline{6-7}
      & & (MeV) & (\%) &  & \multicolumn{1}{c}{Exp.} &
      \multicolumn{1}{c}{Th.} & \\  
      \hline
      $^{41}$Sc($\beta^+){}^{41}$Ca & $7^-, 1$ & 6.496 & 99.963(3) &
      3.461(7) & 2.999 & 4.083 & 6.172 \\ 
      $^{42}$Sc$^*(\beta^+){}^{42}$Ca & $12^+, 2$ & 3.851 & 100 &
      4.17(2) & 2.497 & 3.389  & 11.127 \\ 
      $^{42}$Ti($\beta^+){}^{42}$Sc & $2^+, 0$ & 6.392 & 55(14) &
      3.17(12) & 2.038 & 2.736 &  3.086 \\ 
      $^{43}$Sc($\beta^+){}^{43}$Ca & $7^-, 3$ & 2.221 & 77.5(7) &
      5.03(2) & 0.677 & 0.764 & 6.172 \\ 
      & $5^-, 3$ & 1.848 & 22.5(7) & 4.97(3) & 0.726 & 0.878 & \\ 
      $^{44}$Sc($\beta^+){}^{44}$Ca & $4^+_1,4$ & 2.497 & 98.95(4) &
      5.30(2) & 0.392 & 0.741 & 6.901 \\ 
      & $4^+_2, 4$ & 0.998 & 1.04(4) & 5.15(3) & 0.466 & 0.205 & \\ 
      & $4^+_3, 4$ & 0.353 & 0.010(2) & 6.27(8) & 0.128 & 0.295 & \\ 
      $^{44}$Sc$^*(\beta^+){}^{44}$Ca & $12^+,4$ & 0.640 & 1.20(7) &
      5.88(3) & 0.324 & 0.276 & 11.127 \\ 
      $^{45}$Ca($\beta^-){}^{45}$Sc & $7^-, 3$ & 0.258 & 99.9981 &
      5.983(1) & 0.226 & 0.079 & 13.802 \\ 
      $^{45}$Ti($\beta^+){}^{45}$Sc & $7^-, 3$ & 2.066 & 99.685(17) &
      4.591(2) & 1.123 & 1.551 & 6.172 \\ 
      & $5^-, 3$ & 1.342 & 0.154(12) & 6.24(4) & 0.168 & 0.280 & \\ 
      & $7^-, 3$ & 0.654 & 0.090(10) & 5.81(5) & 0.276 & 0.397 & \\ 
      & $9^-, 3$ & 0.400 & 0.054(5)  & 5.60(4) & 0.351 & 0.712 & \\ 
      $^{45}$V($\beta^+){}^{45}$Ti & $7^-, 1$ & 7.133 & 95.7(15) &
      3.64(2) & 1.801 & 2.208 & 6.172 \\ 
      & $5^-, 1$ & 7.093 & 4.3(15)  & 5.0(2) & 0.701 & 0.428 & \\ 
      $^{46}$Sc$(\beta^-){}^{46}$Ti & $8^+,2$ & 0.357 & 99.9964(7) &
      6.200(3) & 0.187 & 0.277 & 13.093 \\ 
      $^{47}$Ca($\beta^-){}^{47}$Sc & $7^-, 5$ & 1.992 & 19(10) &
      8.5(3) & 0.012 & 0.262 & 16.331 \\ 
      & $5^-, 5$ & 0.695 & 81(10) & 6.04(6) & 0.212 & 0.235 & \\ 
      $^{47}$Sc($\beta^-){}^{47}$Ti & $5^-, 3$ & 0.600 & 31.6(6) &
      6.10(1) & 0.198 & 0.235 & 13.802 \\ 
      & $7^-, 3$ & 0.441 & 68.4(6) & 5.28(1) & 0.508 & 0.611 & 
    \end{tabular}
  \end{center}
\end{table}

\onecolumn
\addtocounter{table}{-1}

\begin{table}
  \begin{center}
    \leavevmode
    \caption{{\em Continuation.\/}}
    \begin{tabular}{lccccrrr}
      Process & $2J_n^\pi, 2T_n^\pi$ & $Q$  & $I_\beta +
      I_\epsilon$ & $\log ft$ & \multicolumn{2}{c}{$M(GT)$} & $W$ \\
      \cline{6-7}
      & & (MeV) & (\%) &  & \multicolumn{1}{c}{Exp.} &
      \multicolumn{1}{c}{Th.} & \\  
      \hline
      $^{47}$V($\beta^+){}^{47}$Ti & $5^-_1, 3$ & 2.928 & 99.552(15) &
      4.901(5) & 0.555 & 0.896 & 4.365 \\ 
      & $3^-_1, 3$ & 1.378 & 0.049(6) & 6.08(6) & 0.143 & 0.107 & \\ 
      & $1^-_1, 3$ & 1.1337 & 0.285(10) & 5.10(2) & 0.442 & 0.563 & \\ 
      & $3^-_2, 3$ & 0.765 & 0.071(3) & 5.36(2) & 0.327 & 0.514 & \\ 
      & $5^-_2, 3$ & 0.761 & 0.0091(7) & 6.25(4) & 0.118 & 0.278 & \\ 
      & $5^-_3, 3$ & 0.402 & 0.0172(9) & 5.41(3) & 0.309 & 0.202 & \\ 
      & $3^-_3, 3$ & 0.379 & 0.0067(5) & 5.77(4) & 0.204 & 0.204 & \\ 
      & $1^-_2, 3$ & 0.134 & 0.0021(6) & 5.18(9) & 0.403 & 0.780 & \\ 
      $^{47}$Cr($\beta^+){}^{47}$V & $3^-, 1$ & 7.451 & 96.1(13) &
      3.70(2) & 0.942 & 1.186 & 4.365 \\ 
      & $5^-, 1$ & 7.363 & 3.9(13) & 5.1(2) & 0.442 & 0.646 & \\
      $^{48}$Sc($\beta^-){}^{48}$Ti & $12^+_1, 4$ & 0.661 & 90.0(3) &
      5.532(13) & 0.484 & 0.780 & 22.256 \\ 
      & $12^+_2, 4$ & 0.485 & 9.85(9) & 6.010(17) & 0.279 & 0.331 & \\ 
      $^{48}$V($\beta^+){}^{48}$Ti & $8^+_1, 4$ & 1.719 & 89.0(9) &
      6.175(7) & 0.192 & 0.345 & 9.259 \\ 
      & $6^+, 4$ & 0.791 & 3.33(7) & 6.565(10) & 0.123 & 0.090 & \\ 
      & $8^+_2, 4$ & 0.775 & 7.76(9) & 6.180(6) & 0.191 & 0.181 & \\ 
      $^{48}$Cr(EC)$^{48}$V & $2^+, 2$ & 1.233 & 100 & 4.294(7) &
      0.559 & 0.709 & 5.346 \\
      $^{48}$Mn($\beta^+){}^{48}$Cr & $8^+_1, 0$ & 11.670 & 6.5(25) &
      5.4(2) & 0.469 & 0.527 & 9.259 \\ 
      & $8^+_2, 0$ & 9.101 & 10.1(24) & 4.6(1) & 1.178 & 2.179 & \\ 
      & $8^+_3, 0$ & 8.876 & 4.0(9) & 5.0(1) & 0.743 & 0.172 & \\ 
      & $8^+_4, 0$ & 8.497 & 8.0(7) & 4.58(5) & 1.206 & 1.361 & \\ 
      & $10^+_1, 0$ & 8.235 & 3.2(4) & 4.90(6) & 0.834 & 0.651 &
    \end{tabular}
  \end{center}
\end{table}

\addtocounter{table}{-1}

\begin{table}
  \begin{center}
    \leavevmode
    \caption{{\em Continuation.\/}}
    \begin{tabular}{lccccrrr}
      Process & $2J_n^\pi, 2T_n^\pi$ & $Q$  & $I_\beta +
      I_\epsilon$ & $\log ft$ & \multicolumn{2}{c}{$M(GT)$} & $W$ \\
      \cline{6-7}
      & & (MeV) & (\%) &  & \multicolumn{1}{c}{Exp.} &
      \multicolumn{1}{c}{Th.} & \\  
      \hline
      $^{49}$Ca($\beta^-){}^{49}$Sc & $3^-, 7$ & 2.178 & 91.5(7) &
      5.075(4) & 0.455 & 1.007 & 13.093 \\ 
      & $5^-_1, 7$ & 1.190 & 7.0(7) & 5.12(5) & 0.432 & 0.209 & \\ 
      & $1^-, 7$ & 0.769 & 0.66(7) & 5.42(5) & 0.306 & 0.757 & \\ 
      & $5^-_2, 7$ & 0.524 & 0.21(6) & 5.3(2) & 0.351 & 0.591 & \\ 
      $^{49}$Sc($\beta^-){}^{49}$Ti & $7^-, 5$ & 1.994 & 99.94(1) &
      5.71(1) & 0.309 & 0.469 & 16.331 \\ 
      & $9^-, 5$ & 0.371 & 0.010(3) & 6.9(2) & 0.079 & 0.072 & \\ 
      & $5^-, 5$ & 0.232 & 0.05(1) & 5.6(1) & 0.351 & 0.389 & \\ 
      $^{49}$V(EC)$^{49}$Ti & $7^-, 5$ & 0.602 & 100 & 6.2(1) & 0.176
      & 0.130 & 10.691 \\ 
      $^{49}$Cr($\beta^+){}^{49}$V & $7^-, 3$ & 2.631 & 12(2) & 5.6(1)
      & 0.304 & 0.335 & 5.346 \\ 
      & $5^-_1, 3$ & 2.540 & 37(2) & 5.02(2) & 0.593 & 0.817 & \\ 
      & $3^-, 3$ & 2.478 & 50(2) & 4.81(2) & 0.755 & 1.033 & \\ 
      & $5^-_2, 3$ & 1.116 & 0.081(9) & 5.80(4) & 0.242 & 0.312 & \\
      & $3^-_2, 3$ & 0.969 & 0.028(6) & 6.15(8) & 0.161 & 0.182 & \\ 
      & $5^-_3, 3$ & 0.396 & 0.0011(2) & 6.75(7) & 0.081 & 0.264 & \\ 
      & $3^-_3, 3$ & 0.322 & 1.9(7) $10^{-4}$ & 7.3(2) & 0.043 & 0.195
      & \\  
      $^{49}$Mn($\beta^+){}^{49}$Cr & $5^-, 1$ & 7.715 & 93.6(26) &
      3.67(3) & 1.364 & 1.704 & 5.346 \\ 
      & $7^-, 1$ & 7.443 & 6.4(26) & 4.8(2) & 0.764 & 0.623 & \\
      $^{50}$Ca($\beta^-){}^{50}$Sc & $2^+, 8$ & 3.118 & 99.0(13) &
      4.14(2) & 0.667 & 0.956 & 6.901 \\
      $^{50}$Sc($\beta^-){}^{50}$Ti & $8^+, 6$ & 4.213 & 8.4(18) &
      6.7(1) & 0.116 & 0.208 & 20.471 \\
      & $12^+, 8$ & 3.689 & 88.4(15) & 5.39(1) & 0.525 & 0.572 & \\
      & $8^+, 8$ & 2.741 & 0.58(4) & 7.01(4) & 0.081 & 0.125 & \\
      & $10^+, 8$ & 2.007 & 1.58(5) & 5.99(2) & 0.263 & 0.358 & \\
%      & $8^+, 8$ & 1.508 & 0.207(21) & 6.37(5) & 0.170 & 0.376 & \\  
    \end{tabular}
  \end{center}
\end{table}

\end{document}